\documentclass[conference]{./pvsctran}

\usepackage{graphicx}
\usepackage{hyperref}
\usepackage{graphicx}
\usepackage{amssymb}
\usepackage{amsmath}
\usepackage[super]{nth}
\usepackage{multicol}
\usepackage{bigints}

\begin{document}
	
	\title{\LARGE Micro-optical Tandem Luminescent Solar Concentrators}
	
	\author{ \IEEEauthorblockN{\large David R. Needell, Ognjen Ilic, Colton R. Bukowsky, Zach Nett, Lu Xu, Junwen He, Haley Bauser, Benjamin G. Lee, John F. Geisz, Ralph G. Nuzzo, A. Paul Alivisatos, and Harry A. Atwater} \\
	\IEEEauthorblockA{\large Institute Name, City, State/Region, Mail/Zip Code, Country (authors' affiliation(s) listed here in 12 point\\ Times font Ð use a second line if necessary)}}

	\author{\IEEEauthorblockN{David R. Needell\IEEEauthorrefmark{1},
			Ognjen Ilic\IEEEauthorrefmark{1}, 
			Colton R. Bukowsky\IEEEauthorrefmark{1},
			Zachary Nett\IEEEauthorrefmark{2},
			Lu Xu\IEEEauthorrefmark{3},
			Junwen He\IEEEauthorrefmark{3},
			Haley Bauser\IEEEauthorrefmark{1},\\
			Benjamin G. Lee\IEEEauthorrefmark{4},
			John F. Geisz\IEEEauthorrefmark{4},
			Ralph G. Nuzzo\IEEEauthorrefmark{3},
			A. Paul Alivisatos\IEEEauthorrefmark{2} and
			Harry A. Atwater\IEEEauthorrefmark{1}}\\
		
		\IEEEauthorblockA{\IEEEauthorrefmark{1}Department of Applied Physics and Materials Science,
			California Institute of Technology,
			Pasadena, CA 91125, USA}
		\IEEEauthorblockA{\IEEEauthorrefmark{2}Department of Chemistry, University of California Berkeley, CA 94720, USA\\}
		\IEEEauthorblockA{\IEEEauthorrefmark{3}Department of Chemistry, University of Illinois at Urbana-Champaign, Urbana, Illinois 61801, USA}
		\IEEEauthorblockA{\IEEEauthorrefmark{4}National Renewable Energy Laboratory, Golden, CO 80401, USA}}
	
	\setlength{\columnsep}{0.25in}
	
	\maketitle
	\begin{abstract}
		Traditional concentrating photovoltaic (CPV) systems utilize multijunction cells to minimize thermalization losses, but cannot efficiently capture diffuse sunlight, which contributes to a high levelized cost of energy (LCOE) and limits their use to geographical regions with high direct sunlight insolation.  Luminescent solar concentrators (LSCs) harness light generated by luminophores embedded in a light-trapping waveguide to concentrate light onto smaller cells.  LSCs can absorb both direct and diffuse sunlight, and thus can operate as flat plate receivers at a fixed tilt and with a conventional module form factor.  However, current LSCs experience significant power loss through parasitic luminophore absorption and incomplete light trapping by the optical waveguide. Here we introduce a tandem LSC device architecture that overcomes both of these limitations, consisting of a PLMA polymer layer with embedded CdSe/CdS quantum dot (QD) luminophores and InGaP micro-cells, which serve as a high bandgap absorber on top of a conventional Si photovoltaic.  We experimentally synthesize CdSe/CdS QDs with exceptionally high quantum-yield (99\%) and ultra-narrowband emission optimally matched to fabricated III-V InGaP micro-cells.  Using a Monte Carlo ray-tracing model, we show the radiative limit power conversion efficiency for a module with these components to be 30.8\% diffuse sunlight conditions.  These results indicate that a tandem LSC-on-Si architecture could significantly improve upon the efficiency of a conventional Si photovoltaic module with simple and straightforward alterations of the module lamination steps of a Si photovoltaic manufacturing process, with promise for widespread module deployment across diverse geographical regions and energy markets.
	\end{abstract}
	
\section*{Introduction}
		
		Silicon photovoltaic (Si PV) modules currently dominate the solar energy market. Progress in Si PV efficiency, combined with historically low module costs, has enabled the overall Levelized Cost of Energy (LCOE) for Si photovoltaics to be competitive with non-renewable energy sources in some markets. However despite recent LCOE reductions, Si PV technology remains economically inferior to fossil fuels\cite{Vest2016}. Additionally, flat-plate Si solar modules require geographical locations with high direct normal incidence (DNI) sunlight conditions in order to maintain module performance \cite{King1997}. Both the strict DNI requirement and the high LCOE of Si PV cells ultimately limits solar power’s dissemination into the global energy market. 
		
		To further PV penetration into electricity markets, luminescent solar concentrators (LSCs) offer a solution to capture diffuse sunlight and reduce the LCOE. A traditional LSC consists of an optical waveguide with luminophores suspended in a polymer matrix and PV material lining the waveguide’s edges \cite{Batchelder1982, Madrid2006, Meinardi2017}. Both diffuse and direct sunlight incident upon this waveguide become absorbed by the embedded luminophores. 
		
		Absorbed photons can undergo radiative recombination, which gives rise to a sharply-peaked and energy down-shifted photoluminesce emission spectrum.  Photons can also undergo nonradiative recombination, becoming parasitically lost as heat. Total internal reflection (TIR) typically guides the photoluminescence radiation to the waveguide edge, where it impinges upon the PV cells \cite{Goetzberger1977, Yablonovitch1980, Yablonovitch1982}. The light concentration factor is directly proportional to the geometric gain (GG) of the LSC –- defined as the ratio of illuminated waveguide area to total PV cell area. 
		
		Despite extensive research and development, LSC module concentration and efficiency suffers from two key loss mechanisms \cite{Gutmann2013, Sark2014}.  First, embedded luminophores require near-unity photoluminescence quantum yield (PLQY) in order to achieve desired optical efficiencies \cite{Martinez2016}.  To prevent excess nonradiative recombination, overlap between luminophore absorbed and emitted photon energies needs to be minimized by employing luminophores that exhibit a Stokes shift \cite{Bronstein2015, Hu2015}.  . Historically, luminophores have not been able to simultaneously exhibit a sufficiently high PLQY and also large Stokes shifts, leading a significant fraction of luminescence photons to be parasitically absorbed by the luminophores. Second, the index of refraction contrast between the optical waveguide and the surrounding medium define a photon escape cone and the limits for waveguide light trapping \cite{Goldschmidt2009}.  Polymer waveguides experience significant escape cone losses for photons radiated at angles that lie between normal incidence and the critical angle of the waveguide. 
		
		\begin{figure*}[t]
			\centering
			\includegraphics[width = \linewidth]{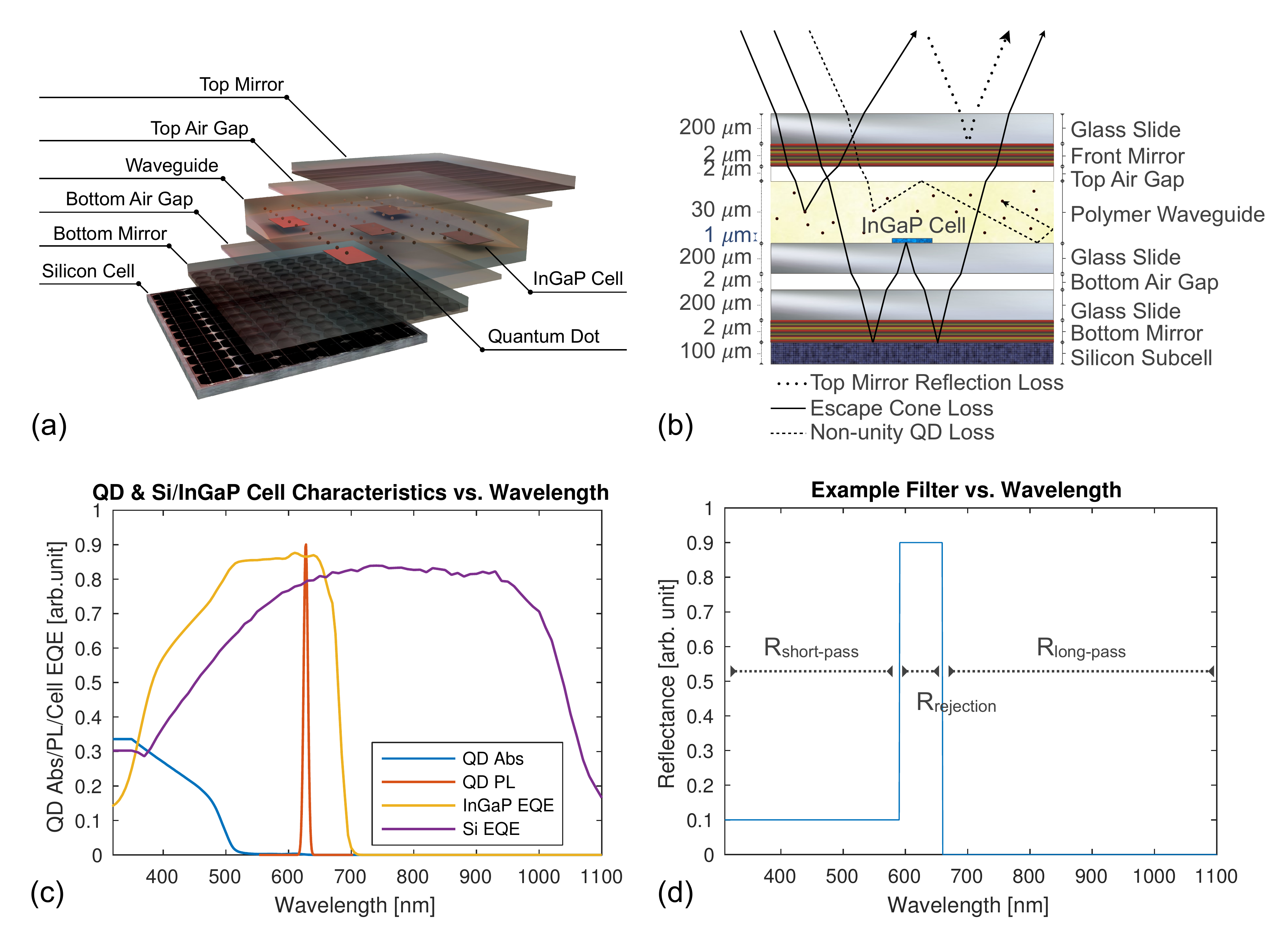}
			\caption{\textit{(a)} 3-dimensional and \textit{(b)} 2-dimensional schematics of the tandem LSC-Si architecture, the QDs are enlarged for viewing purposes, \textit{(c)} measured Si and InGaP EQE curves with respect to wavelength and optimized CdSe/CdS QD absorption and PL spectra; \textit{(d)} an example reflection spectrum dielectric stack filter with respect to wavelength for a given angle of incidence, where $R_\text{short-pass}$, $R_\text{rejection}$, and $R_\text{long-pass}$ correspond to the short-wavelength (300 to 585 nm), mid-wavelength (585 to 665 nm), and long-wavelength (665 to 1100 nm) reflection regimes respectively.}
			\label{Figure: Introduction Diagrams}
		\end{figure*}
		
		Recent advances in cadmium selenide core, cadmium sulfide shell (CdSe/CdS) quantum dot (QD) technology allow for near-unity QDs and sufficiently large Stokes shifts \cite{Bronstein2014, Meinardi2014}.  Furthermore, the addition of top and bottom waveguide coatings that function as spectrally-selective filters presents a possible approach to enhance the waveguide light trapping efficiency. We optically connect a conceptual prototype LSC component in tandem with a planar Si subcell, as shown in Fig. \ref{Figure: Introduction Diagrams}(a) and \ref{Figure: Introduction Diagrams}(b).  Fig. \ref{Figure: Introduction Diagrams}(c) shows the absorption and photoluminescence (PL) spectra for synthesized and measured CdSe/CdS QDs. We further improve upon the performance of conventional LSCs by the use of embedded, planar wide bandgap InGaP micro-cells. By use of a wide bandgap cell whose open circuit voltage exceeds that of a Si cell, LSC-on-Si tandem modules allow greater spectral efficiency across the solar spectrum than Si-only modules.  We can adjust the InGaP cell external quantum efficiency (EQE) spectrum to the PL emission wavelength, as shown in Fig. \ref{Figure: Introduction Diagrams}(c). A planar InGaP cell geometry allows for more rigorous control of the geometric gain, allowing for further optimization of the overall power conversion efficiency (PCE) \cite{Sahin2011}. Last, Fig. \ref{Figure: Introduction Diagrams}(d) shows an ideal filter employing high reflectivity in the rejection band ($R_\text{rejection}$) and low reflectivity in the out of band ($R_\text{short-pass}$, $R_\text{long-pass}$). This LSC design minimizes losses due to reabsorption and nonradiative recombination of luminescence photons and also due to incomplete light trapping. 
		
		Here we characterize a tandem LSC-on-Si module design through the use of a Monte Carlo ray-tracing model. We synthesize and measure CdSe/CdS QDs to obtain experimental absorption/PL data. Additionally, we fabricate InGaP micro-cells and Si subcells to obtain measured EQE as inputs to the Monte Carlo simulation, where cell efficiencies are calculated in the radiative limit. We develop optimal designs for tandem LSC-on-Si modules both with and without spectrally selective mirror layers. Additionally, we optimize device architecture and mirror design for maximum PCE. Finally, we analyze best-case simulation scenarios for the with- and without-mirror cases.

		\begin{figure*}[t]
			\centering
			\includegraphics[width = \linewidth]{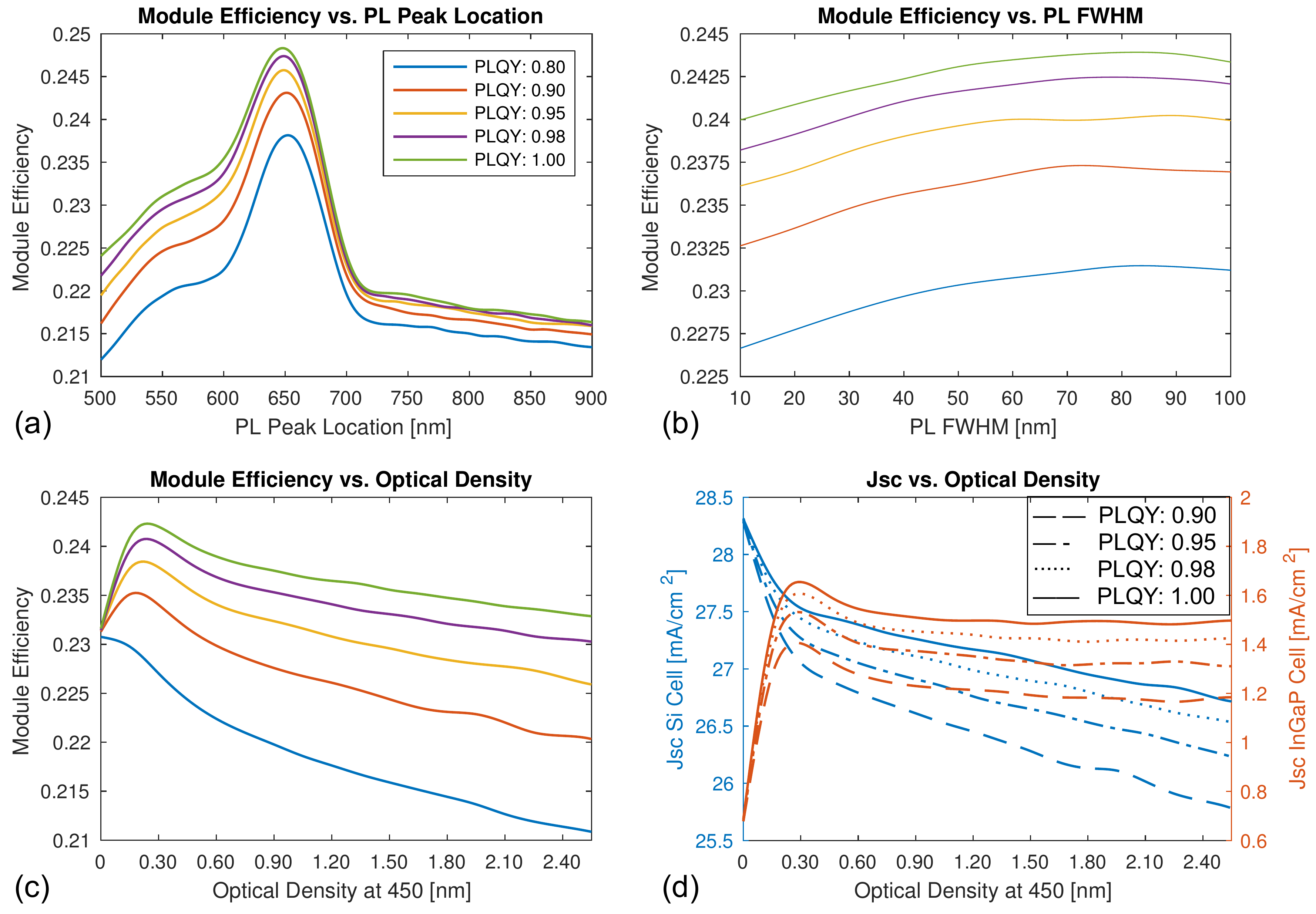}
			\caption{Tandem LSC-Si module efficiency in the case without mirrors, with respect to varying \textit{(a)} PLQY and QD PL peak location assuming an OD of 0.30 and FWHM of 30 nm, \textit{(b)} PLQY and QD PL FWHM assuming an OD of 0.30 and PL peak location of 625 nm, \textit{(c)} PLQY and OD of the embedded QDs at 450 nm assuming a PL peak location of 625 nm and FWHM of 30 nm, and \textit{(d)} short circuit current of the Si and InGaP cells, varying PLQY and OD of the embedded QDs at 450 nm again assuming a PL peak location of 625 nm and FWHM of 30 nm.}
			\label{Figure: No Mirror Optimization}
		\end{figure*}
		
		\section*{Results}
		
		\subsection*{Mirrorless LSC-on-Si Optimization}
		
		The challenge in modeling a tandem LSC-on-Si device arises from the large number of module components. Given the large parameter space of components, we first perform an analysis with extensive multi-parameter variations assuming no top or bottom luminescence photon-trapping mirrors to be present in the device architecture. In our analysis, we varied the QD PL peak spectral position, the QD PL spectral full width at half maximum (FWHM), the optical density (OD) of QDs within the PLMA waveguide, and the QD PLQY. Fig. \ref{Figure: No Mirror Optimization} shows the results of this analysis. 
		
		\begin{figure*}[t]
			\centering
			\includegraphics*[width = \linewidth]{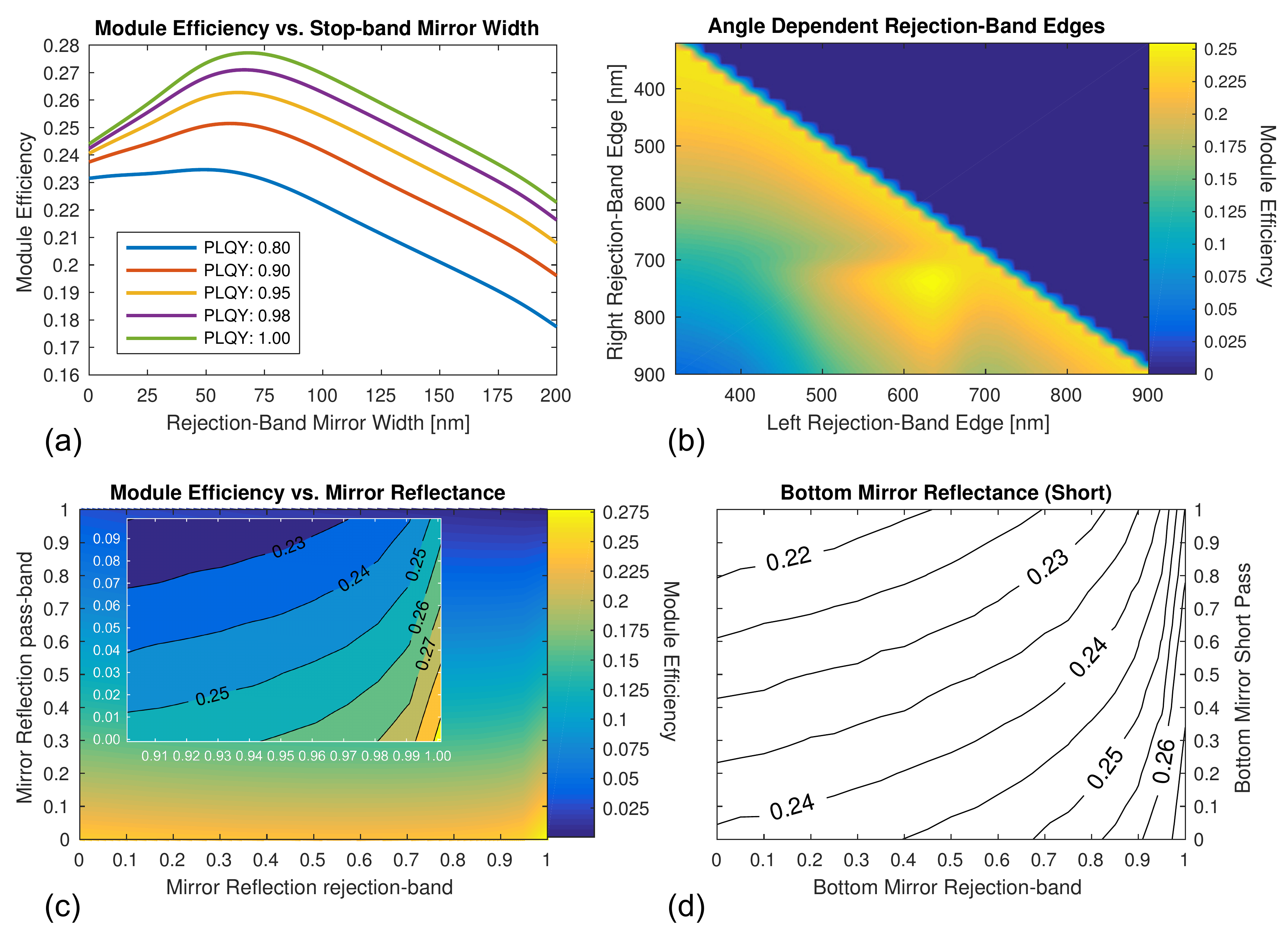}
			\caption{Tandem LSC-Si module efficiency with respect to varying \textit{(a)} rejection-band mirror width, centered at 625 nm, assuming unity rejection-band reflection and unity out-band transmission; \textit{(b)} mirror rejection-band left and right edges' location at DNI photons, assuming unity $R_\text{rejection}$ and zero $R_\text{pass}$; \textit{(c)} top and bottom mirror $R_\text{rejection}$ vs. $R_\text{pass}$ across all angles at ideal rejection-band widths; and \textit{(d)} bottom-only mirror $R_\text{rejection}$ vs. $R_\text{short-pass}$, assuming ideal top mirror performance as shown in (c), and ideal rejection-band widths.}
			\label{Figure: Mirror Optimization}
		\end{figure*}
		
		Without mirrors, the ideal PL peak location gives rise to a maximum module performance at a luminescence wavelength of slightly above 650 nm across all PLQY values, shown in Fig. \ref{Figure: No Mirror Optimization}(a). This results from a Stokes shift increase, minimizing re-absorption losses at wavelengths corresponding to high InGaP cell EQE. We additionally observe an optimum FWHM for the QD PL spectrum of approximately 80 nm, as shown in Fig. \ref{Figure: No Mirror Optimization}(b). For a larger QD PL FWHM, the broader luminescence spectrum yields more frequent opportunities for radiated photons to be absorbed by either the more efficient InGaP micro-cell or the higher EQE Si subcell. In contrast, when the FWHM is too broad, nonradiative recombination via QD re-absorption decreases the overall module efficiency. 
		
		For all assumed PLQY values, we find that maximum PCE occurs at an OD of 0.30 at 450 nm, as shown in Fig. \ref{Figure: No Mirror Optimization}(c). For OD values less than 0.30, QDs do not absorb and then emit enough photons to the InGaP micro-cells. However, for OD values greater than 0.30, photons are either parasitically absorbed by the QDs or reradiated at angles within the escape cone at a greater frequency. We find that with unity PLQY, optimized PL peak location, ideal FWHM, and intermediate waveguide OD we achieve a maximum PCE of $\eta = 24.8\%$ for the mirrorless waveguide design. 
		
		In the mirrorless design, the majority of the output power generated by the tandem LSC-on-Si module comes from the Si subcell. From Fig. \ref{Figure: No Mirror Optimization}(d), we find the generated photocurrent of the Si cell to be an order of magnitude greater than the InGaP cell photocurrent across all PLQY. Even at unity PLQY, escape cone loss prevents a higher concentration of light impinging upon the InGaP cell. We conclude that in order to achieve both higher overall module efficiencies and more significant InGaP cell power generation, additional light-trapping mechanisms must be integrated into the device architecture. 
		
		\subsection*{Top and Bottom Mirror Optimization}
		
		To determine optimal spectral and angular requirements for the top and bottom luminescence photon trapping mirrors, we varied the mirror reflection parameters, with a top hat-like profile, as shown in Fig. \ref{Figure: Introduction Diagrams}(d). The top mirror imposes more stringent spectral requirements, as the optical filter requires near unity transmittance short-pass spectral response, as QD absorption occurs only in this part of the spectrum. Similarly, the mirror long-pass reflectance must be minimized in order for light to reach the Si subcell. In contrast, the bottom mirror short-pass transmitted light is usefully absorbed by the Si subcell, giving rise to power generation, albeit with a less-than-optimal voltage (reduced spectral efficiency). We thus expect the bottom mirror to be less sensitive to short-pass performance than for the top mirror. However, both the top and bottom mirrors require maximum high reflectance in the rejection-band in order to trap luminescence photons impinging on the optical waveguide surface and within the escape cone of the bare waveguide. 
		
		\begin{figure*}[t]
			\centering
			\includegraphics*[width = \linewidth]{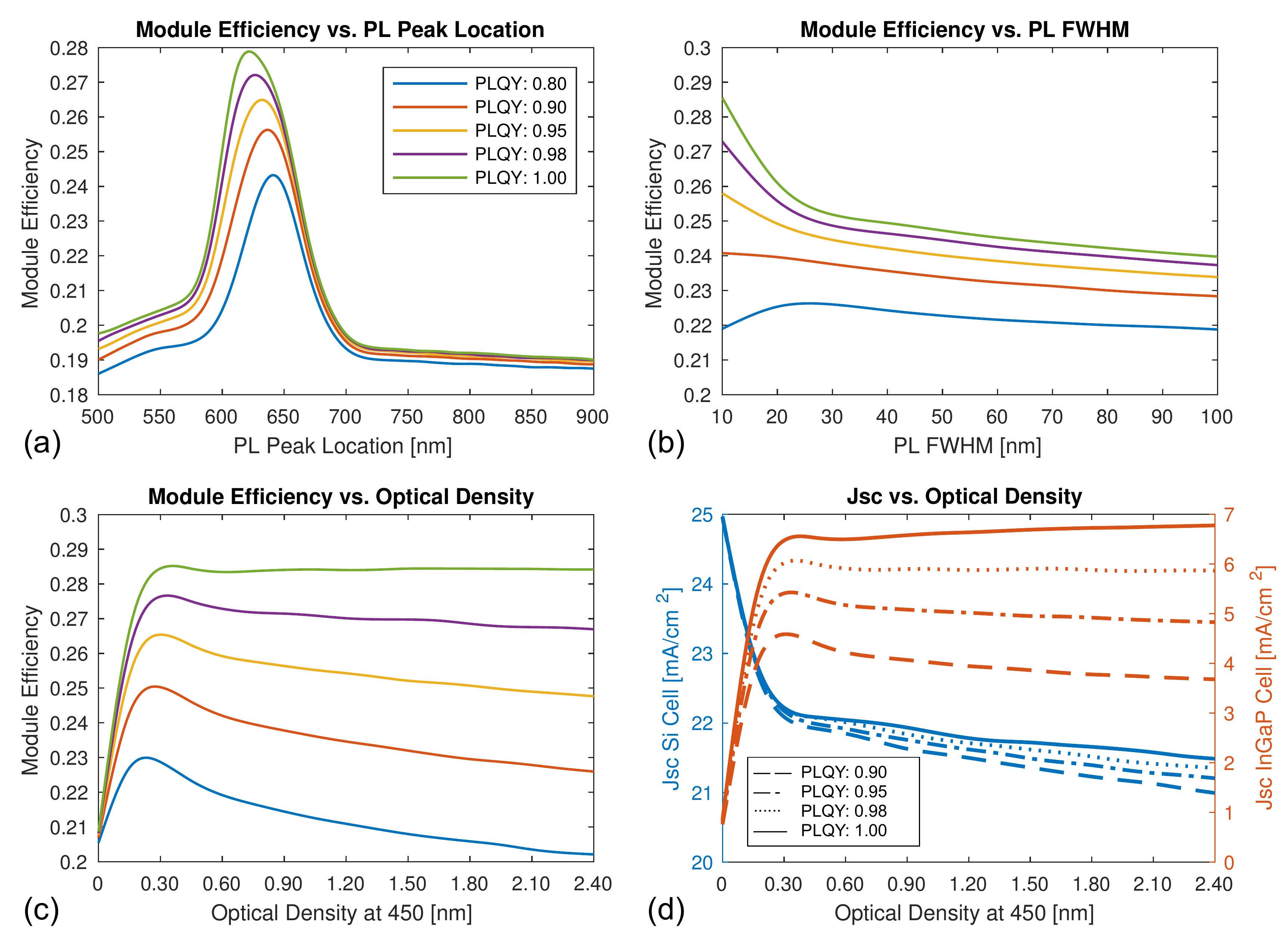}
			\caption{Tandem LSC-Si module efficiency in the perfect-mirrors case with respect to varying \textit{(a)} PLQY and QD PL peak location, \textit{(b)} PLQY and QD PL FWHM, \textit{(c)} PLQY and OD of the embedded QDs at 450 nm, and \textit{(d)} short circuit current of the Si and InGaP cells, varying PLQY and OD of the embedded QDs at 450 nm.}
			\label{Figure: Perfect Mirror Optimization}
		\end{figure*}
		
		Our highest PLQY CdSe/CdS QDs exhibit a FWHM of approximately 30 nm with a PL peak spectral position centered at 625 nm. We therefore treat the FWHM and PL peak spectral position as fixed for purpose of mirror optimization. Given this PL spectrum, Fig. \ref{Figure: Mirror Optimization}(a) shows the module PCE for various rejection-band widths with PL spectral position centered at 625 nm. A FWHM of 68 nm yields maximum device performance, assuming no strong angular dependence of the rejection band. This rejection band mirror reflectance width therefore represents the optimal trade off between capture of QD luminescence photons and parasitic reflection of incident light.  Transmitted photons are absorbed by either the embedded QDs or the underlying Si cell. Assuming ideal rejection-band top/bottom mirror widths, we now investigate the effect of mirror reflectance variation, specifically $R_\text{rejection}$ against $R_\text{pass}$ (in the short-pass and long-pass regimes), where we assume an angular independent response. Fig. \ref{Figure: Mirror Optimization}(c) details the overall module efficiency results, while varying the two reflection parameters, $R_\text{rejection}$ and $R_\text{pass}$. We find that, while optimal module efficiency results from an assumed unity $R_\text{rejection}$ reflectance and zero $R_\text{pass}$, an increase in mirror reflectance $R_\text{pass}$ is more detrimental to overall device performance than a reduction in $R_\text{rejection}$. Assuming no top/bottom mirror parity, we determine the overall impact on module efficiency of varying $R_\text{short-pass}$ for the bottom mirror only, while assuming a unity $R_\text{rejection}$ and zero $R_\text{pass}$ for the top mirror. Fig. \ref{Figure: Mirror Optimization}(d) shows that, as predicted, the short-wavelength transmission requirements for the bottom mirror can be significantly relaxed, still enabling high performance. For the optimal top and bottom angularly independent mirror designs obtained from this analysis, we achieve an overall module efficiency of $\eta = 28.4$\%.
		
		\begin{figure*}[t]
			\centering
			\includegraphics*[width = \linewidth]{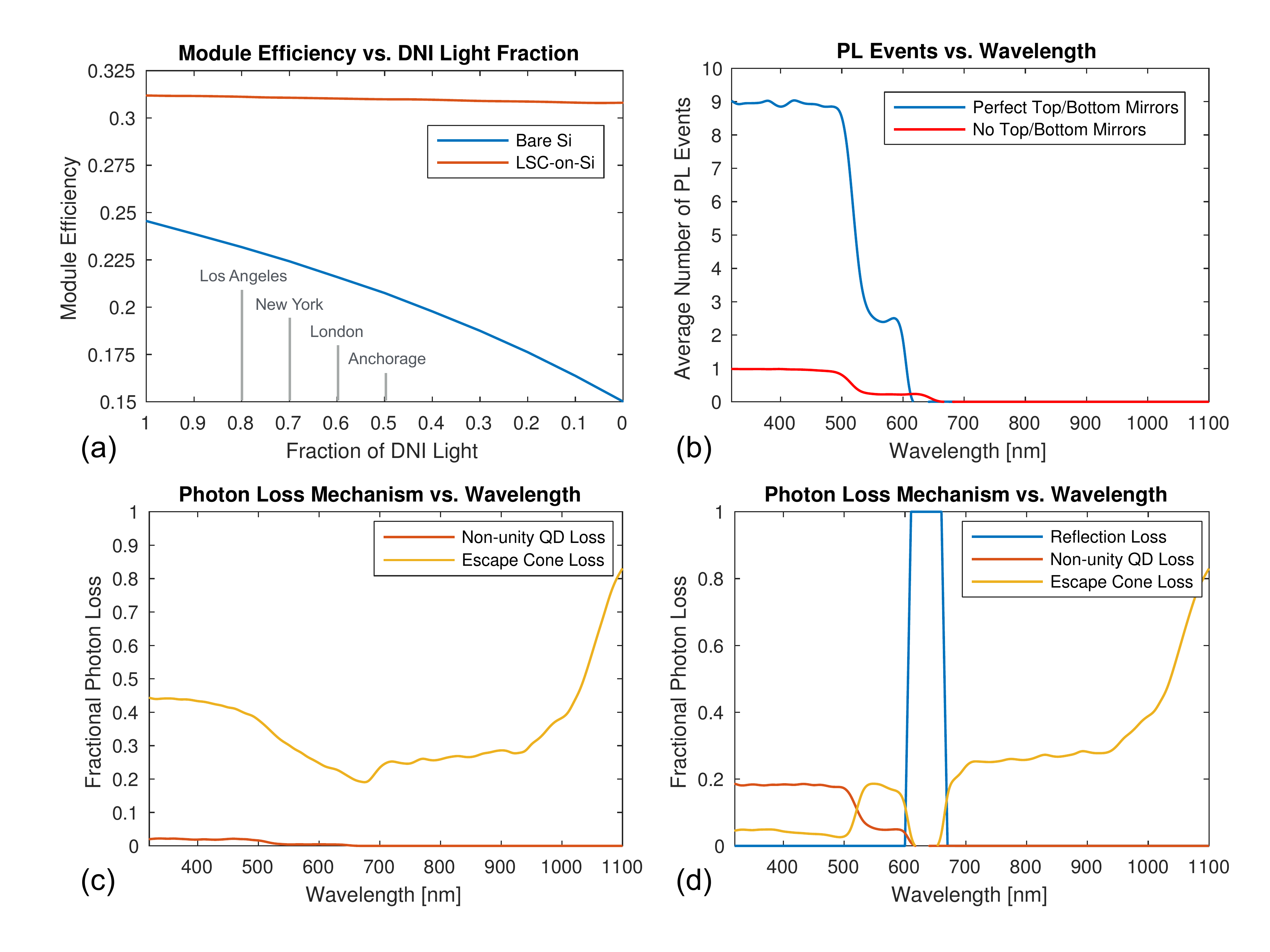}
			\caption{\textit{(a)} A comparison of the tandem LSC-Si module efficiency performance vs. the standalone Si cell efficiency, varying the fraction of light that is normally incident upon the structure; ; \textit{(b)} a comparison between the number of QD absorption and PL events of the cases with/without mirrors under their respective, optimized conditions; and the tandem LSC-Si photon loss mechanisms with respect to incident photon wavelength for the \textit{(c)} case without and \textit{(d)} case with perfect mirrors under their respective, optimized conditions.}
			\label{Figure: Best Case Analysis}
		\end{figure*}
		
		As is the case with common spectrally selective filters (e.g. dielectric stacks), a greater incident photon angle often blueshifts the mirror rejection band while maintaining a relatively constant rejection band width. We simulate varying the rejection-band left and right edges’ location for normally incident photons, and, as an example, assume a rejection-band center shift of 1 nm per degree off normal incidence. Fig. \ref{Figure: Mirror Optimization}(b) shows the results of this simulation. For rejection band left edges located near 620 nm and right edges near 700 nm, we find a globally maximum module efficiency. For rejection-band widths that are too large, module efficiency decreases when the band reflection fails to overlap with the PL spectra and reflects incident light. However, we see that this global maximum occurs at rejection-band widths that, at normal incidence, are further red-shifted. This result implies that the top and bottom mirrors’ fundamental contribution to tandem LSC-on-Si module efficiency comes from reflection of photons with large angles of incidence within the waveguide.
		
		\subsection*{Mirrors LSC-on-Si Optimization}
		
		Analogous to the optimization of the mirrorless tandem LSC, we determine the maximum module efficiency possible with ideal top and bottom mirrors. Fig. \ref{Figure: Perfect Mirror Optimization}(a) shows that ideal PL peak values depend strongly on the PLQY. For lower QD performance, 650 nm yields maximum module efficiency. Parasitic absorption dominates photon recycling events for poor PLQY values. With optimized top/bottom mirrors, photon nonradiative recombination increases for PL peaks centered in the mirrors’ rejection-band. However, as the limit of PLQY approaches unity, the PL peak optimum shifts to 625 nm. Because top/bottom mirror rejection-band reflection is lossless and centered at 625 nm, unity PLQY results in zero instances of nonradiative absorption by the embedded QDs, and therefore allows for unlimited recycling and trapping of short-wavelength photons.
		
		Similarly, we see a strong dependence of module performance on PLQY by varying the FWHM, shown in Fig. \ref{Figure: Perfect Mirror Optimization}(b). We find that, as expected, high PLQY values favor sharply peaked PL. We attribute this to the fact that high PLQY minimizes nonradiative recombination, and therefore narrower FWHM values allow for more photon capture.
		
		Variation of OD of the QDs embedded within the polymer waveguide for the perfect top/bottom mirror case additionally shows a strong dependence on PLQY values, as seen in Fig. \ref{Figure: Perfect Mirror Optimization}(c). However, in all cases , a maximum module PCE is obtained at OD of 0.30 at 450 nm. In contrast to the no-mirror case, we find that the module efficiency decreases less rapidly for increasing OD with non-unity PLQY. With increased QD absorption and PL, there is more opportunity for escape-cone losses. However for ideal mirrors, a unity rejection band redirects all such potential escaped photons back into the waveguide. We therefore find that with unity PLQY, optimized QD PL peak location, ideal FWHM, and an OD of 0.30 at 450 nm, we achieve a maximum PCE of $\eta = 30.8\%$, approximately a 6\% absolute efficiency increase relative to the mirrorless case. 
		
		\subsection*{Optimal Case Analysis}
		
		In the perfect-mirrors case, the output power generated by this tandem LSC-on-Si module is much more evenly split between the Si subcell and the embedded InGaP micro-cell. The InGaP contributes approximately 43\% of the total output power for unity PLQY, as seen in Table 1. Fig. \ref{Figure: Perfect Mirror Optimization}(d) displays the short circuit current contribution of the InGaP cell. In contrast to Fig. \ref{Figure: No Mirror Optimization}(d), the InGaP photocurrent has increased by a factor of 4.5.
		
		\begin{table*}[t]
			\vskip5pt
			\centerline { TABLE 1: \normalsize \textsc{Best Case Simulation Results}}
			\vskip2pt
			\centerline{
				\vbox{\offinterlineskip
					\hrule
					\halign{&\vrule#&
						\strut\quad#\hfil\quad\cr
						height1pt&\omit&&\omit&&\omit&&\omit&&\omit&&\omit&&\omit&\cr
						&{\bf Si Cell Case}&&{\bf Mirror Case}&&PLQY&& Pout (InGaP) [mW/cm$^2$]&&Pout (Si) [mW/cm$^2$]&& Pin [mW/cm$^2$] && {\bf PCE} &\cr
						&\omit&&\omit&&\omit&&\omit&&\omit&&\omit&&\omit&\cr
						\noalign{\hrule}
						&        &&       &&    &&    &&     &&      &&      &\cr
						&Fabricated&&None&&0.98&&2.96&&14.39&&70.2&&24.7&\cr
						&Fabricated&&None&&1.00&&3.03&&14.43&&70.2&&24.8&\cr
						&Fabricated&&Perfect&&0.98&&6.25&&11.40&&70.2 &&25.1&\cr
						&Fabricated&&Perfect&&1.00&&10.21&&11.40&&70.2&&30.8&\cr
						&PERC&&Perfect&&1.00&&10.06&&12.54&&70.2&&32.0&\cr
						&HIT&&Perfect&&1.00&&9.96&&13.17&&70.2&&32.8&\cr}
					\hrule}}
			\label{Table: Best Cases Details}
		\end{table*} 
		
		Fig. \ref{Figure: Best Case Analysis}(b)-(d) compare the performance analysis of the best mirrorless and perfect-mirror designs. For this, we choose non-unity PLQY of 0.98 to determine how parasitic QD absorption loss compares to escape cone loss. We find that in the mirrorless design, the dominant loss comes from escape cone loss, shown in Fig. \ref{Figure: Best Case Analysis}(c). Escape cone loss stems from non-unity Si EQE and incomplete waveguide trapping. For the perfect mirror case, short-wavelength photons are more likely to be parasitically absorbed, as a result of increased photon recycling, shown in Fig. \ref{Figure: Best Case Analysis}(d). We demonstrate this fact by comparing the average number of PL events for a given photon of a certain wavelength, shown in Fig. \ref{Figure: Best Case Analysis}(b). The dominant loss mechanism for long wavelength photons in the perfect mirror design is primarily a result of the imperfect absorption of the Si subcell, matching the mirrorless case. Table 1 shows the comparison between best-case scenarios of the mirrorless and perfect-mirror configuration, for unity and 0.98 PLQY. Additionally, Table 1 compares the overall power output with the use of a 25.6\% cell efficiency, silicon heterojunction structure with interdigitated back contacts as the subcell \cite{Masuko2014}; we also demonstrate the power output possible with this tandem LSC-Si architecture by replacing the subcell with a passivated emitter and rear contact (PERC) Si cell \cite{Padmanabhan2016}.
		
		Finally, we investigate how the fraction of DNI light affects the overall module PCE. Fig. \ref{Figure: Best Case Analysis}(a) shows the results of this simulation for the perfect-mirror architecture versus the standalone Si cell case. As expected, the highest module efficiency results from 100\% DNI for both the LSC-on-Si and bare Si cases; however, we also note that the absolute percent efficiency increase between the completely diffuse case (i.e. 0\% DNI) and the fully direct case is 0.389\% absolute PCE for the LSC-on-Si module.  By contrast, we find that the bare Si cell loses approximately 10\% absolute PCE going from 100\% DNI to 0\%. This suggests that, while DNI light is ideal for maximum module efficiencies, this tandem LSC-on-Si architecture can perform with greater than 30\% PCE even in the completely diffuse limit, and therefore significantly out-compete traditional Si cells in this light insolation regime.  For comparison, D. King \cite{King1997} showed a similar trend of a standalone Si cell’s relative photocurrent with varying angle of incidence.  We understand this result by noting that, even under highly diffuse light, QDs absorb isotropically.  Thus, while the standalone Si cell might reflect light hitting from off-normal incidence angles, a LSC can re-capture this light as it is trapped within the waveguide.  
		
		\section*{Discussion}
		
		We have introduced a tandem-on-Si luminescent solar concentrator.  We show significant PCE enhancements can be achieved in a tandem LSC, relative to both traditional single-layer LSC designs and flat-plate Si cells. A tandem LSC-on-Si module features a number of components that influence its conversion efficiency, including spectrally selective top and bottom mirrors and CdSe/CdS QD luminophores with optimally tuned absorption and PL spectra. We apply a Monte Carlo ray-tracing model to fabricated Si and InGaP cells with measured EQEs, which yield cell efficiencies of $\eta_\text{Si} = 19.8\%$ and $\eta_\text{InGaP} = 26.0\%$ for the radiative limit under low DNI conditions. We ultimately find, under low DNI conditions, maximum LSC-on-Si PCEs reaching 24.8\% and 30.8\% for the mirrorless and perfect-mirror designs respectively. Furthermore, if we assume Si subcell EQEs consistent with reported PERC \cite{Masuko2014} or HIT \cite{Padmanabhan2016} Si cells and perfect mirrors, we find tandem LSC-on-Si PCEs of 32.0\% and 32.8\% could be achieved, respectively.
		
		Assuming a tandem structure without top and bottom mirrors, we find an ideal QD PL peak location of 650 nm given the InGaP and Si cell EQEs, optimized QD PL FWHM of 80 nm, and an OD of the embedded QDs within the PLMA waveguide of 0.30 at 450 nm. For high QD PLQY under these conditions, we find a maximum PCE of 24.8\%, where roughly 17\% of this power is generated by the LSC and 83\% by the Si subcell. 
		
		Optimizing the top and bottom mirrors’ spectral reflection and angular dependence for maximum PCE, we find for near-unity PLQY an ideal QD PL peak location of 625 nm, optimized QD PL FWHM of 10 nm, and an OD of the embedded QDs within the PLMA waveguide of 0.30 at 450 nm. Under these conditions and optimized mirror design, we find a maximum PCE of 30.8\%, where roughly 47\% of the output power is generated by the LSC and 53\% by the Si subcell. Overall, these results indicate the considerable potential for efficiency enhancements by use of a tandem-on-Si LSC architecture, under low DNI conditions. 
	
		\section*{Methods}
		
		The tandem LSC-on-Si module performance is simulated via a Monte Carlo ray-tracing model \cite{Gallagher2004, Papakonstantinou2015, Richards2007}. The algorithm traces photons throughout the module architecture. We determine photon trajectories via scattering, reflection, transmission, and absorption probabilities for each component in the device. We calculate photon reflection probabilities by Fresnel laws for the TE and TM polarizations, and refraction angles via Snell’s law. The algorithm assumes either complete transmission or reflection at a given interface, thereby stochastically treating photon paths \cite{Coropceanu2014, Sholin2007}.  To achieve sufficient statistical averaging, we initialize approximately $10^6$ photons for a given Monte Carlo simulation. To simulate low DNI environments, we assume 40\% of incident photons to normally strike the tandem module, and 60\% to approach with angles uniformly distributed throughout the incident photon hemisphere – i.e. a Lambertian distribution. Cosine factor intensity losses apply to all initialized photons, and determine the net incident power. 
		
		Photons impinge upon either the top filter or the LSC waveguide, for the cases with and without a selective mirror, respectively. We assume a polylaurylmethacrylate  (PLMA) polymer waveguide (refractive index $n = 1.44$) with uniformly distributed QDs. To determine QD absorption within the polymer, we apply the Beer-Lambert law, given a certain optical loading of QDs within the PLMA \cite{Zhou2016}. We input experimentally synthesized CdSe/CdS QDs' absorption and PL characteristics as a baseline for Monte Carlo optimization. 
		
		As mentioned, the heterojunction structure of the CdSe/CdS core/shell QDs allow for fine-tuning of the absorption and photoluminescence spectra. Ideally, luminophores exhibit large Stokes shifts at high PLQY in order to minimize both the number of photons parasitically absorbed by the QDs as well as the amount of light transmitted through the escape cone of the waveguide \cite{Bomm2010}.  To achieve both high Stokes shifts and near-unity PLQY, we follow literature procedures with slight synthesis modifications for developing CdSe/CdS QDs \cite{Chen2013, Jasieniak2009}.  All steady-state absorption spectra are collected using a Shimadzu UV-3600 double beam spectrometer and all steady state PL spectra via a Horiba Jobin-Yvon FluoroLog 2 spectrofluorometer. For the Monte Carlo, we define the PLQY as the probability of photon re-emission directly after absorption by a QD.
		
		In order to account for secondary effects such as polymer non-radiative absorption and photon scattering, we engineer PLMA waveguides of corresponding thicknesses for direct measurement and use in the Monte Carlo. We fabricate luminescent waveguides via the UV polymerization method. We mix the monomer lauryl methacrylate (Sigma Aldrich) and the cross-linker ethylene glycol dimethacrylate (Sigma Aldrich) at a 10:1 volume ratio, and disperse the CdSe/CdS quantum dots in hexane solution. Additionally to control film thickness, we introduce 0.05 vol\% photo-initiator Darocur 1173 (Sigma Aldrich) into solution, where we capillary fill the photo-initiator between two quartz plates with soda lime glass spacers at the corners. Finally, we polymerize the assembled device under UV illumination and inert atmosphere for 30 minutes, followed by removal of the top quartz plate. We measure waveguide absorption using a Varian Cary 5G spectrophotometer.
		
		Once a photon strikes either the embedded InGaP micro-cell or the Si subcell, the measured EQE determines the photon to electron conversion. We fabricate InGaP micro-cells and Si subcells exhibiting EQE shown in Figure \ref{Figure: Introduction Diagrams}(c). We grow upright InGaP solar cells with a thin emitter by metal organic vapor phase epitaxy \cite{Geisz2013} . Micro-cells are processed and placed on glass, using transfer printing techniques \cite{Carlson2012}. The EQE shown in Figure \ref{Figure: Introduction Diagrams}(c) is an angle-averaged EQE calculated for the InGaP device when embedded in PLMA with a 70nm ZnS anti-reflective coating (ARC). We calculate this EQE curve from measurements (and fitting) of larger InGaP devices in air without an ARC. 
		
		For the Si subcell, we use an advanced design suitable for reaching high efficiencies, specifically an interdigitated back passivated contact cell \cite{Efficiency2016}. This back contacted architecture frees the cell of optical shading losses. Passivated contacts enable high open circuit voltages \cite{Feldmann2014, Romer2015}. The cell is fabricated from 180 $\mu$m thick, n-type Cz wafer, with a resistivity of 3 $\Omega$-cm, with wafer saw damage removed by etching and KOH. The wafers are RCA cleaned and a rear tunneling oxide is formed at 700$^\circ$C with O$_2$ flow. We deposit 100 nm of intrinsic amorphous Si on the rear side by PECVD. Beamline ion implantation at 5 kV is done to implant $4\times 10^{15}$ cm$^{-2}$ B for the rear emitter regions, using photolithographic masking to define the area of the device that is implanted. Similarly, we implant, via beamline ion implantation, $7\times 10^{15}$ cm$^{-2}$ P for the rear BSF regions. The sample is annealed at 850$^\circ$C to crystallize the amorphous Si, and activate the implanted dopants. 15 nm of Al$_2$O$_3$ is deposited on the entire surface of the device by ALD and then 75 nm SiNx is deposited on the front side as an anti-reflection coating. The sample is annealed at 400$^\circ$C in forming gas to activate hydrogen passivation of its surfaces. The Al$_2$O$_3$ is etched from the rear of the sample, and a interdigitated pattern is defined photolithographically and then metalized by evaporating 1 $\mu$m Al. We measure the cell EQE using an Oriel QE system. 
		
		To simulate photon reflection via front contact shading, we assign a finite probability to the InGaP cell.  Given our fabricated Si cell EQE, this Monte Carlo simulation yields an overall PCE of $\eta_\text{Si} \approx 21.8\%$ for the stand-alone Si subcell under direct illumination. 
		
		As shown in Figure \ref{Figure: Introduction Diagrams}(b), photon loss mechanisms occur from either initial reflection off of the top mirror of the module, parasitic absorption via the QDs, or transmission through the top surface of the device \cite{Rowan2008}.  A count of the photons and their incident wavelength accepted by either the InGaP or Si cell is integrated with respect to the standard AM1.5G spectrum. The model then uses a detailed balance calculation of the open circuit voltage (V$_\text{oc}$) and fill factor (FF) to give an overall tandem LSC-Si module efficiency \cite{Baruch1995, Levy2006, Rau2014, Shockley1961}.  Note here that we define module efficiency as the generated power ratio to the incident power, and the DNI:diffuse light ratio dictates the incident power. We assume an ideality factor of $n = 1$ for both the InGaP and Si cell cases, and apply the ideal diode equation to determine the V$_\text{oc}$ and FF, given as:
		\begin{equation} \label{Equation: Voc}
			V_\text{oc} = \frac{n k_B T}{q} \ln \left( \frac{I_\text{L}}{I_0} + 1 \right)
		\end{equation}
		where $q$ is the electron charge, $k_B$ the Boltzmann constant, $T$ the cell operating temperature (assumed to be $T=300$K), $I_\text{L}$ is the simulated illumination current, and $I_0$ the dark saturation current.  $I_0$ is approximated based on the radiative limit of the cell, determined from the measured energy bandgaps, $E_g$, of our fabricated InGaP and Si cells:
		\begin{equation}\label{Equation: I0}
			I_0 = \frac{q}{k_B} \frac{15\sigma}{\pi^4} A_\text{wg} T^3 \bigintsss_{E_g/k_BT}^{\infty} \frac{x^2}{e^x - 1} dx
		\end{equation}  
		where $\sigma$ is the Stefan-Boltzmann constant and $A_\text{wg}$ the waveguide area.  This simulation assumes a GG of 100, where we measure the fabricated InGaP micro-cell dimensions to be $1.5 \times 10^{-3}$ m by $1 \times 10^{-4}$ m, yielding an InGaP cell area of 0.15 mm$^2$ and therefore a waveguide area of 15 mm$^2$. 
	
	\section*{Accession Numbers}
	
	For detailed data sets and recorded data, please contact the lead and/or corresponding authors.
	
	\section*{Author Contributions}
	
	Conceptualization, D.R.N., O.I., C.R.B., J.F.G., R.G.N., A.P.A., H.A.A.; Methodology, D.R.N., C.R.B., H.B.; Investigation, D.R.N, Z.N., L.X., J.H., B.G.L., J.F.G.; Software, D.R.N., Visualization, D.R.N.; Writing – Original Draft, D.R.N., O.I., C.R.B.; Writing – Review \& Editing, D.R.N., O.I., C.R.B., H.B., H.A.A., Funding Acquisition, B.G.L., J.F.G., R.G.N., A.P.A., H.A.A.; Resources, H.A.A, A.P.A., R.G.N., J.F.G., B.G.L.
	
	\section*{Acknowledgments}	
	
	This work is made possible by the Advanced Research Projects Agency for Energy (ARPA-E) U.S. Department of Energy grant under the Micro-scale Optimized Solar-cell Arrays with Integrated Concentration (MOSAIC) award, DE- AR0000627.  We thank Noah Bronstein and Sunita Darbe.
	
	\bibliographystyle{IEEEtran}
	
	\bibliography{2017MonteCarloLSCManuscriptReferences}

\begin{thebibliography}{10}
\providecommand{\url}[1]{#1}
\csname url@samestyle\endcsname
\providecommand{\newblock}{\relax}
\providecommand{\bibinfo}[2]{#2}
\providecommand{\BIBentrySTDinterwordspacing}{\spaceskip=0pt\relax}
\providecommand{\BIBentryALTinterwordstretchfactor}{4}
\providecommand{\BIBentryALTinterwordspacing}{\spaceskip=\fontdimen2\font plus
\BIBentryALTinterwordstretchfactor\fontdimen3\font minus
  \fontdimen4\font\relax}
\providecommand{\BIBforeignlanguage}[2]{{%
\expandafter\ifx\csname l@#1\endcsname\relax
\typeout{** WARNING: IEEEtran.bst: No hyphenation pattern has been}%
\typeout{** loaded for the language `#1'. Using the pattern for}%
\typeout{** the default language instead.}%
\else
\language=\csname l@#1\endcsname
\fi
#2}}
\providecommand{\BIBdecl}{\relax}
\BIBdecl

\bibitem{Vest2016}
B.~Vest, ``{Levelized Cost and Levelized Avoided Cost of New Generation
  Resources in the Annual Energy Outlook 2016},'' \emph{Us Eia Lcoe}, no.
  August, pp. 1--20, 2016.

\bibitem{King1997}
D.~King, ``{Photovoltaic Module and Array Performance Characterization Methods
  for All System Operating Conditions},'' \emph{Proceedings of NREL/SNL
  Photovoltaics Program Review Meeting}, pp. 1--22, 1997.

\bibitem{Batchelder1982}
J.~S. Batchelder, ``{The Luminescent Solar Concentrator},'' 1982.

\bibitem{Madrid2006}
\BIBentryALTinterwordspacing
J.~Madrid, M.~Ropp, D.~Galipeau, and S.~May, ``{Investigation of the Efficiency
  Boost Due to Spectral Concentration in a Quantum-Dot Based Luminescent
  Concentrator},'' \emph{2006 IEEE 4th World Conference on Photovoltaic Energy
  Conference}, pp. 154--157, 2006. [Online]. Available:
  \url{http://ieeexplore.ieee.org/lpdocs/epic03/wrapper.htm?arnumber=4059585}
\BIBentrySTDinterwordspacing

\bibitem{Meinardi2017}
\BIBentryALTinterwordspacing
F.~Meinardi, S.~Ehrenberg, L.~Dhamo, F.~Carulli, M.~Mauri, F.~Bruni,
  R.~Simonutti, U.~Kortshagen, and S.~Brovelli, ``{Highly Efficient Luminescent
  Solar Concentrators based on Ultra-Earth-Abundant Indirect Band Gap Silicon
  Quantum Dots},'' \emph{Nature Photonics}, vol.~11, no.~3, pp. 177--185, 2017.
  [Online]. Available: \url{http://dx.doi.org/10.1038/nphoton.2017.5}
\BIBentrySTDinterwordspacing

\bibitem{Goetzberger1977}
a.~Goetzberger and W.~Greubel, ``{Applied Physics Solar Energy Conversion with
  Fluorescent Collectors},'' \emph{Applied Physics}, vol.~14, pp. 123--139,
  1977.

\bibitem{Yablonovitch1980}
E.~Yablonovitch, ``{Thermodynamics of the Fluorescent Planar Concentrator},''
  \emph{Journal of the Optical Society of America}, vol.~70, no.~11, pp.
  1362--1363, 1980.

\bibitem{Yablonovitch1982}
\BIBentryALTinterwordspacing
------, ``{Statistical ray optics},'' \emph{Journal of the Optical Society of
  America}, vol.~72, no.~7, pp. 899--907, 1982. [Online]. Available:
  \url{http://www.opticsinfobase.org/abstract.cfm?URI=josa-72-7-899}
\BIBentrySTDinterwordspacing

\bibitem{Gutmann2013}
J.~Gutmann, H.~Zappe, and J.~C. Goldschmidt, ``{Predicting the performance of
  photonic luminescent solar concentrators},'' \emph{IEEE Journal of
  Photovoltaics}, pp. 1864--1868, 2013.

\bibitem{Sark2014}
W.~G. J. H. M.~V. Sark, Z.~Krumer, C.~D.~M. Doneg{\'{a}}, and R.~E.~I. Schropp,
  ``{Luminescent Solar Concentrators : the route to 10 {\%} efficiency},''
  \emph{IEEE Journal of Photovoltaics}, pp. 2276--2279, 2014.

\bibitem{Martinez2016}
\BIBentryALTinterwordspacing
A.~L. Mart{\'{i}}nez and D.~G{\'{o}}mez, ``{Design, fabrication, and
  characterization of a luminescent solar concentrator with optimized optical
  concentration through minimization of optical losses},'' \emph{Journal of
  Photonics for Energy}, vol.~6, no.~4, p. 045504, 2016. [Online]. Available:
  \url{http://photonicsforenergy.spiedigitallibrary.org/article.aspx?doi=10.1117/1.JPE.6.045504}
\BIBentrySTDinterwordspacing

\bibitem{Bronstein2015}
N.~D. Bronstein, Y.~Yao, L.~Xu, E.~O'Brien, A.~S. Powers, V.~E. Ferry, A.~P.
  Alivisatos, and R.~G. Nuzzo, ``{Quantum Dot Luminescent Concentrator Cavity
  Exhibiting 30-fold Concentration},'' \emph{ACS Photonics}, vol.~2, no.~11,
  pp. 1576--1583, 2015.

\bibitem{Hu2015}
\BIBentryALTinterwordspacing
X.~Hu, R.~Kang, Y.~Zhang, L.~Deng, H.~Zhong, B.~Zou, and L.-J. Shi,
  ``{Ray-trace simulation of CuInS(Se)₂ quantum dot based luminescent solar
  concentrators.}'' \emph{Optics express}, vol.~23, no.~15, pp. A858--67, 2015.
  [Online]. Available:
  \url{http://www.osapublishing.org/viewmedia.cfm?uri=oe-23-15-A858{\&}seq=0{\&}html=true}
\BIBentrySTDinterwordspacing

\bibitem{Goldschmidt2009}
J.~C. Goldschmidt, M.~Peters, A.~B{\"{o}}sch, H.~Helmers, F.~Dimroth, S.~W.
  Glunz, and G.~Willeke, ``{Increasing the efficiency of fluorescent
  concentrator systems},'' \emph{Solar Energy Materials and Solar Cells},
  vol.~93, no.~2, pp. 176--182, 2009.

\bibitem{Bronstein2014}
N.~D. Bronstein, L.~Li, L.~Xu, Y.~Yao, V.~E. Ferry, A.~P. Alivisatos, and R.~G.
  Nuzzo, ``{Luminescent solar concentration with semiconductor nanorods and
  transfer-printed micro-silicon solar cells},'' \emph{ACS Nano}, vol.~8,
  no.~1, pp. 44--53, 2014.

\bibitem{Meinardi2014}
\BIBentryALTinterwordspacing
F.~Meinardi, A.~Colombo, K.~A. Velizhanin, R.~Simonutti, M.~Lorenzon,
  L.~Beverina, R.~Viswanatha, V.~I. Klimov, and S.~Brovelli, ``{Large-area
  luminescent solar concentrators based on /`Stokes-shift-engineered/'
  nanocrystals in a mass-polymerized PMMA matrix},'' \emph{Nat Photon}, vol.~8,
  no.~5, pp. 392--399, 2014. [Online]. Available:
  \url{http://dx.doi.org/10.1038/nphoton.2014.54}
\BIBentrySTDinterwordspacing

\bibitem{Sahin2011}
\BIBentryALTinterwordspacing
D.~Şahin, B.~Ilan, and D.~F. Kelley, ``{Monte-Carlo simulations of light
  propagation in luminescent solar concentrators based on semiconductor
  nanoparticles},'' \emph{Journal of Applied Physics}, vol. 110, no.~3, p.
  033108, 2011. [Online]. Available:
  \url{http://scitation.aip.org/content/aip/journal/jap/110/3/10.1063/1.3619809}
\BIBentrySTDinterwordspacing

\bibitem{Gallagher2004}
\BIBentryALTinterwordspacing
S.~J. Gallagher, P.~C. Eames, and B.~Norton, ``{Quantum dot solar concentrator
  behaviour, predicted using a ray trace approach},'' \emph{Int. J. Ambient
  Energy}, vol.~25, no.~1, pp. 47--56, 2004. [Online]. Available:
  \url{http://dx.doi.org/10.1080/01430750.2004.9674937}
\BIBentrySTDinterwordspacing

\bibitem{Papakonstantinou2015}
\BIBentryALTinterwordspacing
I.~Papakonstantinou and C.~Tummeltshammer, ``{Fundamental limits of
  concentration in luminescent solar concentrators revised: the effect of
  reabsorption and nonunity quantum yield},'' \emph{Optica}, vol.~2, no.~10, p.
  841, 2015. [Online]. Available:
  \url{http://www.osapublishing.org/viewmedia.cfm?uri=optica-2-10-841{\&}seq=0{\&}html=true}
\BIBentrySTDinterwordspacing

\bibitem{Richards2007}
B.~S. Richards and K.~R. McIntosh, ``{Overcoming the poor short wavelength
  spectral response of CdS/CdTe photovoltaic modules via luminescence
  down-shifting: Ray-tracing simulations},'' \emph{Progress in Photovoltaics:
  Research and Applications}, vol.~15, no.~1, pp. 27--34, 2007.

\bibitem{Coropceanu2014}
I.~Coropceanu and M.~G. Bawendi, ``{Core/shell quantum dot based luminescent
  solar concentrators with reduced reabsorption and enhanced efficiency},''
  \emph{Nano Letters}, vol.~14, no.~7, pp. 4097--4101, 2014.

\bibitem{Sholin2007}
V.~Sholin, J.~D. Olson, and S.~A. Carter, ``{Semiconducting polymers and
  quantum dots in luminescent solar concentrators for solar energy
  harvesting},'' \emph{Journal of Applied Physics}, vol. 101, no.~12, 2007.

\bibitem{Zhou2016}
Y.~Zhou, D.~Benetti, Z.~Fan, H.~Zhao, D.~Ma, A.~O. Govorov, A.~Vomiero, and
  F.~Rosei, ``{Near Infrared, Highly Efficient Luminescent Solar
  Concentrators},'' \emph{Advanced Energy Materials}, vol.~6, no.~11, pp. 1--8,
  2016.

\bibitem{Bomm2010}
J.~Bomm, A.~B{\"{u}}chtemann, A.~Fiore, L.~Manna, J.~H. Nelson, D.~Hill, and
  W.~G. J. H.~M. van Sark, ``{Fabrication and spectroscopic studies on highly
  luminescent CdSe/CdS nanorod polymer composites},'' \emph{Beilstein Journal
  of Nanotechnology}, vol.~1, no.~1, pp. 94--100, 2010.

\bibitem{Chen2013}
\BIBentryALTinterwordspacing
O.~Chen, J.~Zhao, V.~P. Chauhan, J.~Cui, C.~Wong, D.~K. Harris, H.~Wei, H.-S.
  Han, D.~Fukumura, R.~K. Jain, and M.~G. Bawendi, ``{Compact high-quality
  CdSe-CdS core-shell nanocrystals with narrow emission linewidths and
  suppressed blinking.}'' \emph{Nature materials}, vol.~12, no.~5, pp. 445--51,
  2013. [Online]. Available:
  \url{http://www.pubmedcentral.nih.gov/articlerender.fcgi?artid=3677691{\&}tool=pmcentrez{\&}rendertype=abstract}
\BIBentrySTDinterwordspacing

\bibitem{Jasieniak2009}
J.~Jasieniak, L.~Smith, J.~van Embden, P.~Mulvaney, and M.~Califano,
  ``{Re-examination of the size dependent absorption properties of CdSe quantum
  dots},'' \emph{Journal of Physical Chemistry C}, vol. 113, pp.
  19\,468--19\,474, 2009.

\bibitem{Geisz2013}
J.~F. Geisz, M.~A. Steiner, I.~Garc{\'{i}}a, S.~R. Kurtz, and D.~J. Friedman,
  ``{Enhanced external radiative efficiency for 20.8{\%} efficient
  single-junction GaInP solar cells},'' \emph{Applied Physics Letters}, vol.
  103, no.~4, 2013.

\bibitem{Carlson2012}
A.~Carlson, A.~M. Bowen, Y.~Huang, R.~G. Nuzzo, and J.~A. Rogers, ``{Transfer
  printing techniques for materials assembly and micro/nanodevice
  fabrication},'' \emph{Advanced Materials}, vol.~24, no.~39, pp. 5284--5318,
  2012.

\bibitem{Efficiency2016}
S.~Essig, M.~A. Steiner, C.~Alleb, J.~F. Geisz, B.~Paviet-salomon, S.~Ward,
  A.~Descoeudres, V.~Lasalvia, L.~Barraud, N.~Badel, A.~Faes, J.~Levrat,
  M.~Despeisse, C.~Ballif, P.~Stradins, and D.~L. Young, ``{Realization of
  GaInP / Si Dual-Junction Solar Cells with 29.8{\%} 1-Sun Efficiency},''
  \emph{IEEE Journal of Photovoltaics}, vol.~6, no.~4, pp. 1012--1019, 2016.

\bibitem{Feldmann2014}
\BIBentryALTinterwordspacing
F.~Feldmann, M.~Simon, M.~Bivour, C.~Reichel, M.~Hermle, and S.~W. Glunz,
  ``{Carrier-selective contacts for Si solar cells},'' \emph{Applied Physics
  Letters}, vol. 104, no.~18, p. 181105, 2014. [Online]. Available:
  \url{http://scitation.aip.org/content/aip/journal/apl/104/18/10.1063/1.4875904}
\BIBentrySTDinterwordspacing

\bibitem{Romer2015}
U.~R{\"{o}}mer, R.~Peibst, T.~Ohrdes, B.~Lim, J.~Kr{\"{u}}gener, T.~Wietler,
  and R.~Brendel, ``{Ion implantation for poly-Si passivated back-junction
  back-contacted solar cells},'' \emph{IEEE Journal of Photovoltaics}, vol.~5,
  no.~2, pp. 507--514, 2015.

\bibitem{Rowan2008}
B.~C. Rowan, L.~R. Wilson, and B.~S. Richards, ``{Advanced material concepts
  for luminescent solar concentrators},'' \emph{IEEE Journal on Selected Topics
  in Quantum Electronics}, vol.~14, no.~5, pp. 1312--1322, 2008.

\bibitem{Baruch1995}
\BIBentryALTinterwordspacing
P.~Baruch, A.~{De Vos}, P.~Landsberg, and J.~Parrott, ``{On some thermodynamic
  aspects of photovoltaic solar energy conversion},'' \emph{Solar Energy
  Materials and Solar Cells}, vol.~36, no.~2, pp. 201--222, 1995. [Online].
  Available: \url{http://linkinghub.elsevier.com/retrieve/pii/0927024895800042}
\BIBentrySTDinterwordspacing

\bibitem{Levy2006}
M.~Y. Levy and C.~Honsberg, ``{Rapid and precise calculations of energy and
  particle flux for detailed-balance photovoltaic applications},''
  \emph{Solid-State Electronics}, vol.~50, no. 7-8, pp. 1400--1405, 2006.

\bibitem{Rau2014}
U.~Rau, U.~W. Paetzold, and T.~Kirchartz, ``{Thermodynamics of light management
  in photovoltaic devices},'' \emph{Physical Review B - Condensed Matter and
  Materials Physics}, vol.~90, 2014.

\bibitem{Shockley1961}
W.~Shockley and H.~J. Queisser, ``{Detailed balance limit of efficiency of p-n
  junction solar cells},'' \emph{Journal of Applied Physics}, vol.~32, no.~3,
  pp. 510--519, 1961.

\bibitem{Masuko2014}
K.~Masuko, M.~Shigematsu, T.~Hashiguchi, D.~Fujishima, M.~Kai, N.~Yoshimura,
  T.~Yamaguchi, Y.~Ichihashi, T.~Mishima, N.~Matsubara, T.~Yamanishi,
  T.~Takahama, M.~Taguchi, E.~Maruyama, and S.~Okamoto, ``{Achievement of More
  Than 25{\%} Conversion Efficiency With Crystalline Silicon Heterojunction
  Solar Cell},'' \emph{IEEE Journal of Photovoltaics}, vol.~4, no.~6, pp.
  1433--1435, 2014.

\bibitem{Padmanabhan2016}
M.~Padmanabhan, K.~Jhaveri, R.~Sharma, P.~K. Basu, S.~Raj, J.~Wong, and J.~Li,
  ``{Light-induced degradation and regeneration of multicrystalline silicon
  Al-BSF and PERC solar cells},'' \emph{Physica Status Solidi - Rapid Research
  Letters}, vol.~10, no.~12, pp. 874--881, 2016.

\end{thebibliography}
	
\end{document}